\documentclass[showpacs]{revtex4}
\usepackage{indentfirst}
\begin{document}
\title{\bf Low-lying hypernuclei in the relativistic quark-gluon model}
\author{S.M. Gerasyuta}
\email{gerasyuta@SG6488.spb.edu}
\author{E.E. Matskevich}
\email{matskev@pobox.spbu.ru}
\affiliation{Department of Theoretical Physics, St. Petersburg State University, 198904,
St. Petersburg, Russia}
\affiliation{Department of Physics, LTA, 194021, St. Petersburg, Russia}
\begin{abstract}
Low-lying hypernuclei $ ^3_{\Lambda}H$, $ ^3_{\Sigma}H$, $ ^3_{\Lambda}He$,
$ ^3_{\Sigma}He$ are described by the relativistic nine-quark equations
in the framework of the dispersion relation technique. The approximate
solutions of these equations using the method based on the extraction
of leading singularities of the amplitudes are obtained. The relativistic
nine-quark amplitudes of hypernuclei, including the quarks of three
flavors ($u$, $d$, $s$) are calculated. The poles of these amplitudes
determine the masses of hypernuclei. The mass of state $ ^3_{\Lambda}H$
with the isospin $I=0$ and the spin-parity $J^P=\frac{1}{2}^+$ is
equal to $M=2991\, MeV$.
\end{abstract}
\maketitle
\section{Introduction.}
Hypernuclei spectroscopy is enjoying an experimental renaissance with
ongoing and planned program at DA$\Phi$NE, FAIR, Jefferson Lab, J-PARC
and Mainz providing motivation for enhanced theoretical efforts
(for a recent review, see Ref. \cite{1}).

LQCD calculations \cite{2, 3, 4, 5, 6} are performed using an isotropic
quark action at the $SU(3)$ flavor symmetric point corresponding to the
physical strange quark mass, with $m_{\pi}=m_K=m_{\eta}\sim 800\, MeV$.
The lattice spacing $b=0.145\, fm$ has been used in the calculations,
dictated by the avaible computational resources, and therefore an
extrapolation to the continuum has not been performed. Further, extrapolation
of the physical pion mass have not been attempted, because the quark mass
dependence of the energy levels in the light nuclei are not known. Future
calculations at smaller lattice spacing and at lighter quark masses will
facilitate such extrapolations and lead to first predictions for the
spectrum of light nuclei, with completely quantified uncertainties, that
can be compared with experiment.

In the recent paper \cite{7} the relativistic six-quark equations are found
in the framework of coupled-channel formalism. The dynamical mixing between
the subamplitudes of hexaquark are considered. The six-quark amplitudes
of dibaryons are calculated. The poles of these amplitudes determine the
masses of dibaryons. We calculated the contribution of six-quark
subamplitudes to the hexaquark amplitudes.

In the previous paper \cite{8} the $ ^3 He$ as the system of interacting quarks
and gluons is considered. The relativistic nine-quark equations are found
in the framework of the dispersion relation technique. The dynamical
mixing between the subamplitudes of $ ^3 He$ is taken into account.
The relativistic nine-quark amplitudes of $ ^3 He$, including the
$u$, $d$ quarks are calculated. The approximate solutions
of these equations using the method based on the extraction of leading
singularities of the amplitudes were obtained. The poles of the nonaquark
amplitudes determined the mass of $ ^3 He$.

The experimental mass value of $ ^3 He$ is equal to $M=2808.392413\, MeV$.
The experimental data of hypertriton mass is equal to $M=2991.166905\, MeV$.
The model use only three parameters which are determined by these masses:
the cutoff $\Lambda=9$ and the gluon coupling constant $g=0.2122$.
The mass of $u$-quark is equal $m=410\, MeV$, the mass of strange quark $m_s=607\, MeV$
takes into account the confinement potential (shift mass is equal to $50\, MeV$).

In the present paper the hypernuclei $ ^3_{\Lambda}H$, $ ^3_{\Sigma}H$,
$ ^3_{\Lambda}He$, $ ^3_{\Sigma}He$, $nn \Lambda$ and $nn \Sigma$ are
considered in the framework of the dispersion relation technique.
The approximate solutions of the nine-quark equations using the method
based on the extraction of leading singularities of the amplitude are
obtained. The relativistic nona-amplitudes of low-lying hypernuclei,
including the three flavors ($u$, $d$, $s$) are calculated. The poles
of these amplitudes determine the masses of the hypernuclei.

In Sec. II the relativistic nine-quark amplitudes of the hypernuclei
are constructed. The dynamical mixing between the subamplitudes of
hypernuclei are considered. Sec. III is devoted to the calculation
results for the masses of the lowest hypernuclei (Table \ref{tab1}).
In conclusion, the status of the considered model is discussed.

\section{Nine-quark amplitudes of hypernuclei.}

We derive the relativistic nine-quark equations in the framework of the
dispersion relation technique. We use only planar diagrams; the other
diagrams due to the rules of $1/N_c$ expansion \cite{9, 10, 11} are neglected.

The current generates a nine-quark system. The correct equations for the
amplitude are obtained by taking into account all possible subamplitudes.
Then one should represent a nine-particle amplitude as a sum of 36 subamplitudes:

\begin{eqnarray}
\label{1}
A=\sum\limits_{i<j \atop i, j=1}^9 A_{ij}\, . \end{eqnarray}

This defines the division of the diagrams into groups according to the
certain pair interaction of particles. The total amplitude can be
represented graphically as a sum of diagrams. We need to consider only
one group of diagrams and the amplitude corresponding to them, for example
$A_{12}$. We consider the derivation of the relativistic generalization
of the Faddeev-Yakubovsky approach \cite{12, 13}.

In our case the two states of hypertritons are obtained: with the isospin
$I=0$ $ ^3_{\Lambda} H$ ($uud\, udd\, uds$) and with the isospin $I=1$
$ ^3_{\Sigma} H$ ($uud\, udd\, uus$). The similar method allows us to
construct the hyperhelium states: with the isospin $I=0$ $ ^3_{\Sigma} He$
($uud\, uud\, dds$), with the isospin $I=1$ $ ^3_{\Lambda} He$ ($uud\, uud\, uds$)
and with the isospin $I=2$ $ ^3_{\Sigma} He$ ($uud\, uud\, uus$).

We take into account the pairwise interaction of all nine quarks.
The set of diagrams associated with the amplitude $A_{12}$ can further be
broken down into some groups corresponding to subamplitudes similar to \cite{8},
Appendix A.

In order to represent the subamplitudes $A_i$ in the form of a dispersion relation,
it is necessary to define the amplitude of $qq$ interactions. We use the results
of our relativistic quark model \cite{14}:

\begin{equation}
\label{2}
a_n(s_{ik})=\frac{G^2_n(s_{ik})}
{1-B_n(s_{ik})} \, ,\end{equation}

\begin{equation}
\label{3}
B_n(s_{ik})=\int\limits_{(m_i+m_k)^2}^{\Lambda\frac{(m_i+m_k)^2}{4}}
\, \frac{ds'_{ik}}{\pi}\frac{\rho_n(s'_{ik})G^2_n(s'_{ik})}
{s'_{ik}-s_{ik}} \, .\end{equation}

\noindent
Here $G_n(s_{ik})$ are the diquark vertex functions (Table \ref{tab2}).
The vertex functions are determined by the contribution of the crossing
channels. The vertex functions satisfy the Fierz relations. These vertex
functions are generated from $g_V$. $B_n(s_{ik})$ and $\rho_n (s_{ik})$
are the Chew-Mandelstam functions with cutoff $\Lambda$ and the phase spaces:

\begin{eqnarray}
\label{4}
\rho_n (s_{ik},J^P)&=&\left(\alpha(n,J^P) \frac{s_{ik}}{(m_i+m_k)^2}
+\beta(n,J^P)+\delta(n,J^P) \frac{(m_i-m_k)^2}{s_{ik}}\right)
\nonumber\\
&&\nonumber\\
&\times & \frac{\sqrt{(s_{ik}-(m_i+m_k)^2)(s_{ik}-(m_i-m_k)^2)}}
{s_{ik}}\, .
\end{eqnarray}

The coefficients $\alpha(n,J^P)$, $\beta(n,J^P)$ and
$\delta(n,J^P)$ are given in Table \ref{tab2}.

Here $n=1$ coresponds to $qq$-pairs with $J^P=0^+$, $n=2$ corresponds
to the $qq$ pairs with $J^P=1^+$. $s_{ik}$ is the two-particle subenergy
squared.

\begin{table}
\caption{Masses of lowest hypernuclei. Parameters of model: $\Lambda=9.0$,
$g=0.2122$, $m=410\, MeV$, $m_s=607\, MeV$.}\label{tab1}
\begin{tabular}{|c|c|c|c|c|}
\hline
\, $I$ \, & \quad\quad $J^P$ \quad\quad & \, hypernuclei \, & \, quark content \, & \, mass, MeV \,  \\
\hline
0 & $\frac{1}{2}^+$, $\frac{3}{2}^+$ & $ ^3_{\Lambda} H$ & $uud \, udd \, uds$ & 2991 \\ [6pt]
\hline
1 & $\frac{1}{2}^+$, $\frac{3}{2}^+$ & $ ^3_{\Sigma} H$ & $uud \, udd \, uus$ & 2993 \\
&&& $(uud \, udd \, dds)$ & \\
\hline
0 & $\frac{1}{2}^+$, $\frac{3}{2}^+$ & $ ^3_{\Sigma} He$ & $uud \, uud \, dds$ & 3017 \\ [6pt]
1 & $\frac{1}{2}^+$, $\frac{3}{2}^+$ & $ ^3_{\Lambda} He$ & $uud \, uud \, uds$ & 2991 \\ [6pt]
2 & $\frac{1}{2}^+$, $\frac{3}{2}^+$ & $ ^3_{\Sigma} He$ & $uud \, uud \, uus$ & 2953 \\ [6pt]
\hline
\end{tabular}
\end{table}

In the case in question the interacting quarks do not produce a bound
states, therefore the integration is carried out from the threshold
$(m_i+m_k)^2$ to the cutoff $\Lambda$.

Let us extract singularities in the coupled equations and obtain
the reduced amplitudes $\alpha_i$.

The isospin $I=0$ $ ^3_{\Lambda}H$.

We need to consider the system of 29 equations:

\begin{eqnarray}
\label{5}
\alpha_1: & & \, \, 1^{uu}, \, 1^{dd}, \, 0^{ud}, \, 0^{us}, \, 0^{ds}
\\
\label{6}
\alpha_2: & & \, \, 1^{uu}1^{uu}, \, 1^{uu}1^{dd}, \, 1^{uu}0^{ud},
 \, 1^{uu}0^{us}, \, 1^{uu}0^{ds}, \nonumber
\\
 & & \, \, 1^{dd}1^{dd}, \, 1^{dd}0^{ud}, \, 1^{dd}0^{us}, \, 1^{dd}0^{ds}, \nonumber
\\
 & & \, \, 0^{ud}0^{ud}, \, 0^{ud}0^{us}, \, 0^{ud}0^{ds}
 \\
\label{7}
\alpha_3: & & \, \, 1^{uu}1^{dd}0^{ud}, \, 1^{uu}1^{dd}0^{us}, \, 1^{uu}1^{dd}0^{ds},
 \nonumber
\\
 & & \, \, 1^{uu}0^{ud}0^{ud}, \, 1^{uu}0^{ud}0^{us}, \, 1^{uu}0^{ud}0^{ds}, \nonumber
\\
 & & \, \, 0^{ud}1^{dd}0^{ud}, \, 0^{ud}1^{dd}0^{us}, \, 0^{ud}1^{dd}0^{ds}, \nonumber
\\
 & & \, \, 0^{ud}0^{ud}0^{ud}, \, 0^{ud}0^{ud}0^{us}, \, 0^{ud}0^{ud}0^{ds}
\end{eqnarray}

The isospin $I=1$ $ ^3_{\Sigma}H$ (24 equations):

\begin{eqnarray}
\label{8}
\alpha_1: & & \, \, 1^{uu}, \, 1^{dd}, \, 0^{ud}, \, 0^{us}, \, 0^{ds}
\\
\label{9}
\alpha_2: & & \, \, 1^{uu}1^{uu}, \, 1^{uu}1^{dd}, \, 1^{uu}0^{ud},
 \, 1^{uu}0^{us}, \, 1^{uu}0^{ds}, \nonumber
\\
 & & \, \, 1^{dd}0^{ud}, \, 1^{dd}0^{us}, \, 1^{dd}0^{ds}, \nonumber
\\
 & & \, \, 0^{ud}0^{ud}, \, 0^{ud}0^{us}, \, 0^{ud}0^{ds}
 \\
\label{10}
\alpha_3: & & \, \, 1^{uu}1^{dd}1^{uu}, \, 1^{uu}1^{dd}0^{us},
 \nonumber
\\
 & & \, \, 1^{uu}0^{ud}1^{uu}, \, 1^{uu}0^{ud}0^{us}, \nonumber
\\
 & & \, \, 0^{ud}1^{dd}1^{uu}, \, 0^{ud}1^{dd}0^{us}, \nonumber
\\
 & & \, \, 0^{ud}0^{ud}1^{uu}, \, 0^{ud}0^{ud}0^{us}
\end{eqnarray}

Here the $\alpha_1$ are determined by the diquarks, the $\alpha_2$
includes the two diquarks and the five quarks. $\alpha_3$ corresponds
to the $pp\Lambda$ ($pn\Lambda$) states.

The equations for the $ ^3_{\Lambda}He$, $ ^3_{\Sigma}He$ are considered.

The isospin $I=0$.

We need to consider 23 equations:

\begin{eqnarray}
\label{11}
\alpha_1: & & \, \, 1^{uu}, \, 1^{dd}, \, 0^{ud}, \, 0^{us}, \, 0^{ds}
\\
\label{12}
\alpha_2: & & \, \, 1^{uu}1^{uu}, \, 1^{uu}1^{dd}, \, 1^{uu}0^{ud},
 \, 1^{uu}0^{us}, \, 1^{uu}0^{ds}, \nonumber
\\
 & & \, \, 1^{dd}1^{dd}, \, 1^{dd}0^{ud}, \, 1^{dd}0^{us}, \, 1^{dd}0^{ds}, \nonumber
\\
 & & \, \, 0^{ud}0^{ud}, \, 0^{ud}0^{us}, \, 0^{ud}0^{ds}
 \\
\label{13}
\alpha_3: & & \, \, 1^{uu}1^{uu}1^{dd}, \, 1^{uu}1^{uu}0^{ds},
 \nonumber
\\
 & & \, \, 1^{uu}0^{ud}1^{dd}, \, 1^{uu}0^{ud}0^{ds}, \nonumber
\\
 & & \, \, 0^{ud}0^{ud}1^{dd}, \, 0^{ud}0^{ud}0^{ds}
\end{eqnarray}

The isospin $I=1$ (25 equations):

\begin{eqnarray}
\label{14}
\alpha_1: & & \, \, 1^{uu}, \, 1^{dd}, \, 0^{ud}, \, 0^{us}, \, 0^{ds}
\\
\label{15}
\alpha_2: & & \, \, 1^{uu}1^{uu}, \, 1^{uu}1^{dd}, \, 1^{uu}0^{ud},
 \, 1^{uu}0^{us}, \, 1^{uu}0^{ds}, \nonumber
\\
 & & \, \, 1^{dd}0^{ud}, \, 1^{dd}0^{us}, \, 1^{dd}0^{ds}, \nonumber
\\
 & & \, \, 0^{ud}0^{ud}, \, 0^{ud}0^{us}, \, 0^{ud}0^{ds}
 \\
\label{16}
\alpha_3: & & \, \, 1^{uu}1^{uu}0^{ud}, \, 1^{uu}1^{uu}0^{us}, \, 1^{uu}1^{uu}0^{ds},
 \nonumber
\\
 & & \, \, 1^{uu}0^{ud}0^{ud}, \, 1^{uu}0^{ud}0^{us}, \, 1^{uu}0^{ud}0^{ds}, \nonumber
\\
 & & \, \, 0^{ud}0^{ud}0^{ud}, \, 0^{ud}0^{ud}0^{us}, \, 0^{ud}0^{ud}0^{ds}
\end{eqnarray}

The isospin $I=2$ (20 equations):

\begin{eqnarray}
\label{17}
\alpha_1: & & \, \, 1^{uu}, \, 1^{dd}, \, 0^{ud}, \, 0^{us}, \, 0^{ds}
\\
\label{18}
\alpha_2: & & \, \, 1^{uu}1^{uu}, \, 1^{uu}1^{dd}, \, 1^{uu}0^{ud},
 \, 1^{uu}0^{us}, \, 1^{uu}0^{ds}, \nonumber
\\
 & & \, \, 1^{dd}0^{us}, \nonumber
\\
 & & \, \, 0^{ud}0^{ud}, \, 0^{ud}0^{us}, \, 0^{ud}0^{ds}
 \\
\label{19}
\alpha_3: & & \, \, 1^{uu}1^{uu}1^{uu}, \, 1^{uu}1^{uu}0^{us},
 \nonumber
\\
 & & \, \, 1^{uu}0^{ud}1^{uu}, \, 1^{uu}0^{ud}0^{us}, \nonumber
\\
 & & \, \, 0^{ud}0^{ud}1^{uu}, \, 0^{ud}0^{ud}0^{us}
\end{eqnarray}

The coefficients of the coupled equations are determined by the
permutation of quarks \cite{12, 13}. For the simplicity we
consider the graphical equation of the reduced amplitude
$\alpha_3^{1^{uu}1^{uu}1^{uu}}$ (Fig. 1).

In Fig. 1 the first coefficient is equal to 12, that the number $12=2$
(permutation of particles 1 and 2) $\times 2$ (permutation of particles
3 and 4) $\times 3$ (permutation of pairs (56) and (12), (56) and (34));
the second coefficient equal to 12, that the number $12=2$
(permutation of particles 1 and 2) $\times 3$ (permutation of pairs
(12) and (34), (12) and (56)) $\times 2$ (we can replace 7-th $d$-quark
with 8-th $d$-quark); the third coefficient equal to 6, that the number
$6=2$ (permutation of particles 1 and 2) $\times 3$ (permutation of pairs
(12) and (34), (12) and (56)); the 4-th coefficient equal to 6: $6=2$
(permutation of $d$-quarks 7 and 8) $\times 3$ (permutation of pairs
(12) and (34), (12) and (56)); the 5-th coefficient equal to 12: $12=2$
(permutation of $d$-quarks 7 and 9) $\times 2$ (permutation of $d$-quarks
7 and 8) $\times 3$ (permutation of pairs (12) and (34), (12) and (56));
the 6-th coefficient equal to 24: $24=2$ (permutation of particles 1 and 2)
$\times 2$ (permutation of particles 3 and 4) $\times 3$ (permutation of
pairs (56) and (12), (56) and (34)) $\times 2$ (permutation of $d$-quarks
7 and 8); the 7-th coefficient equal to 48: $48=2$ (permutation of particles
1 and 2) $\times 2$ (permutation of particles 3 and 4) $\times 3$ (permutation
of pairs (56) and (12), (56) and (34)) $\times 2$ (permutation of $d$-quarks
7 and 9) $\times 2$ (permutation of $d$-quark 7 and $s$-quark 8); the 8-th
coefficient equal to 48: $48=2$ (permutation of particles 1 and 2) $\times 2$
(permutation of particles 3 and 4) $\times 2$ (permutation of pairs (12) and
(34)) $\times 3$ (permutation of pairs (56) and (12), (56) and (34)) $\times 2$
(permutation of $d$-quarks 7 and 8); the 9-th coefficient equal to 24: $24=2$
(permutation of particles 1 and 2) $\times 2$ (permutation of particles 3 and 4)
$\times 2$ (permutation of pairs (12) and (34)) $\times 3$ (permutation of pairs
(56) and (12), (56) and (34)); the 10-th coefficient equal to 24 (this case is
similar to 6-th); the 11-th coefficient equal to 48 (this case is similar
to 7-th); the 12-th coefficient equal to 48: $48=2$ (permutation of particles
1 and 2) $\times 6$ (permutation of pairs (12), (34) and (56)) $\times 2$
(permutation of $d$-quarks 7 and 8) $\times 2$ (permutation of $d$-quark 8 and
$s$-quark 9); the 13-th coefficient equal to 24: $24=2$ (permutation of particles
1 and 2) $\times 6$ (permutation of pairs (12), (34) and (56)) $\times 2$
(permutation of $d$-quarks 8 and 9).

\begin{table}
\caption{The vertex functions and coefficients of Chew-Mandelstam functions.}\label{tab2}
\begin{tabular}{|c|c|c|c|c|c|}
\hline
\, $n$ \, & \, $J^P$ \, & $G_n^2(s_{kl})$ & \, $\alpha_n$ \, & $\beta_n$ & \, $\delta_n$ \, \\
\hline
& & & & & \\
1 & $0^+$ & $\frac{4g}{3}-\frac{8g m_{kl}^2}{(3s_{kl})}$
& $\frac{1}{2}$ & $-\frac{1}{2}\frac{(m_k-m_l)^2}{(m_k+m_l)^2}$ & $0$ \\
& & & & & \\
2 & $1^+$ & $\frac{2g}{3}$ & $\frac{1}{3}$
& $\frac{4m_k m_l}{3(m_k+m_l)^2}-\frac{1}{6}$
& $-\frac{1}{6}$ \\
& & & & & \\
\hline
\end{tabular}
\end{table}

The similar approach allows us to take into account the coefficients
in all equations.

The functions $I_1$ -- $I_{10}$ are taken from the paper \cite{7}.
The other functions are determined by the following formulae \cite{8}
(see also Fig. 1):

\begin{eqnarray}
\label{20}
I_{11}(12,34,15,36,47)=I_1(12,15)\times I_2(34,36,47) \, ,
\end{eqnarray}

\begin{eqnarray}
\label{21}
I_{12}(12,34,56,17)=I_1(12,17) \, ,
\end{eqnarray}

\begin{eqnarray}
\label{22}
I_{13}(12,34,56,17,28)=I_2(12,17,28) \, ,
\end{eqnarray}

\begin{eqnarray}
\label{23}
I_{14}(12,34,56,17,38)=I_1(12,17)\times I_1(34,38) \, ,
\end{eqnarray}

\begin{eqnarray}
\label{24}
I_{15}(12,34,56,23,47)=I_7(12,34,23,47) \, ,
\end{eqnarray}

\begin{eqnarray}
\label{25}
I_{16}(12,34,56,17,23,48)=I_8(12,34,17,23,48) \, ,
\end{eqnarray}

\begin{eqnarray}
\label{26}
I_{17}(12,34,56,17,45)=I_1(12,17)\times I_3(34,56,45) \, ,
\end{eqnarray}

\begin{eqnarray}
\label{27}
I_{18}(12,34,56,17,38,49)=I_1(12,17)\times I_2(34,38,49) \, ,
\end{eqnarray}

\begin{eqnarray}
\label{28}
I_{19}(12,34,56,17,38,59)=I_1(12,17)\times I_1(34,38)\times I_1(56,59) \, ,
\end{eqnarray}

\begin{eqnarray}
\label{29}
I_{20}(12,34,56,17,45,68)=I_1(12,17)\times I_7(34,56,45,68) \, ,
\end{eqnarray}

\begin{eqnarray}
\label{30}
I_{21}(12,34,56,17,28,45)=I_2(12,17,28)\times I_3(34,56,45) \, .
\end{eqnarray}

The main contributions are determined by the functions $I_1$ and $I_2$:

\begin{eqnarray}
\label{31}
I_1(ij)&=&\frac{B_j(s_0^{13})}{B_i(s_0^{12})}
\int\limits_{(m_1+m_2)^2}^{\Lambda\frac{(m_1+m_2)^2}{4}}
\frac{ds'_{12}}{\pi}\frac{G_i^2(s_0^{12})\rho_i(s'_{12})}
{s'_{12}-s_0^{12}} \int\limits_{-1}^{+1} \frac{dz_1(1)}{2}
\frac{1}{1-B_j (s'_{13})}\, , \\
&&\nonumber\\
\label{32}
I_2(ijk)&=&\frac{B_j(s_0^{13}) B_k(s_0^{24})}{B_i(s_0^{12})}
\int\limits_{(m_1+m_2)^2}^{\Lambda\frac{(m_1+m_2)^2}{4}}
\frac{ds'_{12}}{\pi}\frac{G_i^2(s_0^{12})\rho_i(s'_{12})}
{s'_{12}-s_0^{12}}
\frac{1}{2\pi}\int\limits_{-1}^{+1}\frac{dz_1(2)}{2}
\int\limits_{-1}^{+1} \frac{dz_2(2)}{2}\nonumber\\
&&\nonumber\\
&\times&
\int\limits_{z_3(2)^-}^{z_3(2)^+} dz_3(2)
\frac{1}{\sqrt{1-z_1^2(2)-z_2^2(2)-z_3^2(2)+2z_1(2) z_2(2) z_3(2)}}
\nonumber\\
&&\nonumber\\
&\times& \frac{1}{1-B_j (s'_{13})} \frac{1}{1-B_k (s'_{24})}
 \, ,
\end{eqnarray}

\noindent
where $i$, $j$, $k$ correspond to the diquarks with the
spin-parity $J^P=0^+$, $1^+$.

The one new contribution $I_{22}\sim10^{-8}$. The contributions
of $I_{10}$, $I_{17}$, $I_{19}$, $I_{20}$, $I_{21}$ also are
small $\sim10^{-7}$. We do not take into account these functions
in the coupled equations.

\section{Calculation results and conclusions.}

The poles of the reduced amplitudes correspond to the bound state
and determine the masses of low-lying hypernuclei. The quark masses
$m=410\, MeV$ and $m_s=607\, MeV$ coincide with the quark masses of
the ordinary baryons \cite{15, 16}. The calculated mass values of the
hypernuclei are shown in the Table \ref{tab1}.

We predict the degeneracy of hypertriton and hyperhelium with the
spin-parity $J^P=\frac{1}{2}^+$, $\frac{3}{2}^+$.

The hypertriton with the $I=0$ $ ^3_{\Lambda} H$ and the hyperhelium
with the $I=1$ $ ^3_{\Lambda} He$ also can be considered with the degeneracy.
The estimation of theoretical error is equal to $1\, MeV$.

It is now clear, but hardly a surprise, that the spectrum of nuclei and
hypernuclei changes dramatically from light-quark masses. While one had
already shown this from the recent work \cite{17} on the $H$-dibaryon,
and nucleon-nucleon scattering lengths, this has now been demonstrated
to the true for even larger systems.

It will be interesting to learn how the various thresholds for binding
evolve with the light quark masses providing accurate binding energies
for any given light quark masses will require the inclusion of
electromagnetic effect.

Lattice QCD has evolved to the point where first principles calculation
of light nuclei are now possible, as demonstrated by the calculation at
unphysically heavy quark masses presented in the paper \cite{18}. The
experimental program in hypernuclear physics, and the difficulties
encountered in accurately determined rates for low energy nuclear
reactions, warrant continued effort in, and development of, the
application of LQCD to nuclear physics.

Cleary, calculation at smaller lattice spacings at $SU(3)$ symmetric
point are required in order to remove the systematic uncertainties in
the nuclear binding energies at these quark masses. In order to impact
directly the experimental program in nuclear and hypernuclear physics,
analogous calculations must be performed at lighter quark masses,
ideally at or close to their physical values.

The antihypertriton have the mass that is similar to the hypertriton
in our calculations.

\begin{acknowledgments}
The authors would like to thank T. Barnes and L.V. Krasnov for useful discussions.
\end{acknowledgments}

\newpage

\begin{picture}(600,100)
\put(-20,54){\line(1,0){18}}
\put(-20,53){\line(1,0){18}}
\put(-20,52){\line(1,0){18}}
\put(-20,51){\line(1,0){18}}
\put(-20,50){\line(1,0){18}}
\put(-20,49){\line(1,0){18}}
\put(-20,48){\line(1,0){18}}
\put(-20,47){\line(1,0){18}}
\put(-20,46){\line(1,0){18}}
\put(10,50){\circle{25}}
\put(-1,46){\line(1,1){15}}
\put(2,41){\line(1,1){17}}
\put(7.5,38.5){\line(1,1){14}}
\put(6,61.5){\vector(1,3){10}}
\put(19,85){7}
\put(6,39){\vector(1,-3){10}}
\put(19,10){8}
\put(1,42){\vector(1,-4){8.5}}
\put(0,9){9}
\put(20,68){\circle{16}}
\put(20,32){\circle{16}}
\put(31,50){\circle{16}}
\put(28,70){\vector(3,2){19}}
\put(50,80){1}
\put(28,70){\vector(3,-1){22}}
\put(53,63){2}
\put(39,50){\vector(3,2){14}}
\put(55,52){3}
\put(39,50){\vector(3,-2){14}}
\put(55,41){4}
\put(28,30){\vector(3,1){22}}
\put(53,30){5}
\put(28,30){\vector(3,-2){19}}
\put(50,13){6}
\put(14,65){\small $1^{uu}$}
\put(14,29){\small $1^{uu}$}
\put(25,47){\small $1^{uu}$}
\put(60,84){$u$}
\put(63,64){$u$}
\put(63,54){$u$}
\put(63,41){$u$}
\put(63,30){$u$}
\put(60,11){$u$}
\put(27,88){$d$}
\put(26,2){$d$}
\put(10,-2){$s$}
\put(74,48){=}
\put(-5,-25){$\alpha_3^{1^{uu}1^{uu}1^{uu}}$}
\put(87,54){\line(1,0){19}}
\put(87,53){\line(1,0){20}}
\put(87,52){\line(1,0){21}}
\put(87,51){\line(1,0){22}}
\put(87,50){\line(1,0){23}}
\put(87,49){\line(1,0){22}}
\put(87,48){\line(1,0){21}}
\put(87,47){\line(1,0){20}}
\put(87,46){\line(1,0){19}}
\put(105,55){\vector(1,4){8.2}}
\put(117,86){7}
\put(105,45){\vector(1,-4){8.2}}
\put(116,08){8}
\put(105,45){\vector(0,-1){34}}
\put(97,10){9}
\put(121,67){\circle{16}}
\put(127,50){\circle{16}}
\put(121,32){\circle{16}}
\put(129,70){\vector(1,1){13}}
\put(131,82){1}
\put(129,70){\vector(3,-1){18}}
\put(148,67){2}
\put(135,50){\vector(3,2){16}}
\put(155,55){3}
\put(135,50){\vector(3,-2){16}}
\put(155,39){4}
\put(129,30){\vector(3,1){18}}
\put(148,26){5}
\put(129,30){\vector(1,-1){13}}
\put(132,9){6}
\put(115,64){\small $1^{uu}$}
\put(121,47){\small $1^{uu}$}
\put(115,29){\small $1^{uu}$}
\put(140,89){$u$}
\put(158,70){$u$}
\put(164,60){$u$}
\put(164,37){$u$}
\put(158,24){$u$}
\put(145,8){$u$}
\put(125,92){$d$}
\put(122,-2){$d$}
\put(105,-2){$s$}
\put(178,48){+}
\put(115,-25){$\lambda$}
\put(197,54){\line(1,0){18}}
\put(197,53){\line(1,0){18}}
\put(197,52){\line(1,0){18}}
\put(197,51){\line(1,0){18}}
\put(197,50){\line(1,0){18}}
\put(197,49){\line(1,0){18}}
\put(197,48){\line(1,0){18}}
\put(197,47){\line(1,0){18}}
\put(197,46){\line(1,0){18}}
\put(227,50){\circle{25}}
\put(216,46){\line(1,1){15}}
\put(219,41){\line(1,1){17}}
\put(224.5,38.5){\line(1,1){14}}
\put(227,30){\circle{16}}
\put(227,22){\vector(2,-3){11}}
\put(240,10){6}
\put(227,22){\vector(-2,-3){11}}
\put(209,10){5}
\put(247,50){\circle{16}}
\put(255,50){\vector(3,2){17}}
\put(262,48){2}
\put(255,50){\vector(3,-2){17}}
\put(268,43){3}
\put(232,61){\vector(1,0){40}}
\put(255,66){1}
\put(232,39){\vector(1,0){40}}
\put(255,28){4}
\put(230,62){\vector(1,4){7}}
\put(239,85){7}
\put(227,62){\vector(0,1){29}}
\put(229,87){8}
\put(224,62){\vector(-1,4){7}}
\put(220,87){9}
\put(280,60){\circle{16}}
\put(280,40){\circle{16}}
\put(288,61){\vector(3,1){20}}
\put(313,70){1}
\put(288,61){\vector(3,-1){20}}
\put(313,53){2}
\put(288,39){\vector(3,1){20}}
\put(313,41){3}
\put(288,39){\vector(3,-1){20}}
\put(313,24){4}
\put(221,27){\small $1^{uu}$}
\put(241,47){\small $1^{uu}$}
\put(274,57){\small $1^{uu}$}
\put(274,37){\small $1^{uu}$}
\put(321,74){$u$}
\put(321,53){$u$}
\put(321,43){$u$}
\put(321,21){$u$}
\put(207,-1){$u$}
\put(243,-1){$u$}
\put(243,97){$d$}
\put(231,99){$d$}
\put(216,97){$s$}
\put(263,65){$u$}
\put(255,55){$u$}
\put(255,41){$u$}
\put(263,30){$u$}
\put(332,48){+}
\put(220,-25){$12\, I_{9}(1^{uu}1^{uu}1^{uu}1^{uu})\, \alpha_1^{1^{uu}}$}
\put(347,54){\line(1,0){18}}
\put(347,53){\line(1,0){18}}
\put(347,52){\line(1,0){18}}
\put(347,51){\line(1,0){18}}
\put(347,50){\line(1,0){18}}
\put(347,49){\line(1,0){18}}
\put(347,48){\line(1,0){18}}
\put(347,47){\line(1,0){18}}
\put(347,46){\line(1,0){18}}
\put(377,50){\circle{25}}
\put(366,46){\line(1,1){15}}
\put(369,41){\line(1,1){17}}
\put(374.5,38.5){\line(1,1){14}}
\put(390,66){\circle{16}}
\put(396,71){\vector(1,1){16}}
\put(403,87){7}
\put(396,71){\vector(2,-1){18.5}}
\put(406,68){1}
\put(389,53){\vector(3,1){25.5}}
\put(403,50){2}
\put(423,62){\circle{16}}
\put(431,62){\vector(2,1){20}}
\put(457,68){1}
\put(431,62){\vector(2,-1){20}}
\put(457,50){2}
\put(376,62){\vector(1,2){16}}
\put(394,85){8}
\put(371,60.5){\vector(1,4){9}}
\put(382,90){9}
\put(394,39){\circle{16}}
\put(372,30.5){\circle{16}}
\put(400.5,34){\vector(4,-1){21}}
\put(424,25){3}
\put(400.5,34){\vector(1,-4){5.2}}
\put(408,7){4}
\put(371,22.5){\vector(2,-3){12}}
\put(386,3){5}
\put(371,22.5){\vector(-2,-3){12}}
\put(353,3){6}
\put(384,63){\small $0^{ud}$}
\put(417,59){\small $1^{uu}$}
\put(388,36){\small $1^{uu}$}
\put(366,27.5){\small $1^{uu}$}
\put(465,73){$u$}
\put(465,47){$u$}
\put(432,20){$u$}
\put(402,-1){$u$}
\put(386,-8){$u$}
\put(352,-8){$u$}
\put(414,93){$d$}
\put(394,100){$d$}
\put(381,103){$s$}
\put(412,73){$u$}
\put(411,49){$u$}
\put(357,-25){$12\, I_{12}(1^{uu}1^{uu}1^{uu}0^{ud})\, \alpha_1^{0^{ud}}$}
\end{picture}

\vskip110pt

\begin{picture}(600,60)
\put(10,48){+}
\put(25,54){\line(1,0){18}}
\put(25,53){\line(1,0){18}}
\put(25,52){\line(1,0){18}}
\put(25,51){\line(1,0){18}}
\put(25,50){\line(1,0){18}}
\put(25,49){\line(1,0){18}}
\put(25,48){\line(1,0){18}}
\put(25,47){\line(1,0){18}}
\put(25,46){\line(1,0){18}}
\put(55,50){\circle{25}}
\put(44,46){\line(1,1){15}}
\put(47,41){\line(1,1){17}}
\put(52.5,38.5){\line(1,1){14}}
\put(68,66){\circle{16}}
\put(74,71){\vector(1,1){16}}
\put(81,87){7}
\put(74,71){\vector(2,-1){18.5}}
\put(84,68){1}
\put(67,53){\vector(3,1){25.5}}
\put(81,50){2}
\put(101,62){\circle{16}}
\put(109,62){\vector(2,1){20}}
\put(135,68){1}
\put(109,62){\vector(2,-1){20}}
\put(135,50){2}
\put(54,62){\vector(1,2){16}}
\put(72,85){8}
\put(49,60.5){\vector(1,4){9}}
\put(60,90){9}
\put(72,39){\circle{16}}
\put(50,30.5){\circle{16}}
\put(78.5,34){\vector(4,-1){21}}
\put(102,25){3}
\put(78.5,34){\vector(1,-4){5.2}}
\put(86,7){4}
\put(49,22.5){\vector(2,-3){12}}
\put(64,3){5}
\put(49,22.5){\vector(-2,-3){12}}
\put(31,3){6}
\put(62,63){\small $0^{us}$}
\put(95,59){\small $1^{uu}$}
\put(66,36){\small $1^{uu}$}
\put(44,27.5){\small $1^{uu}$}
\put(143,73){$u$}
\put(143,47){$u$}
\put(110,20){$u$}
\put(80,-1){$u$}
\put(64,-8){$u$}
\put(30,-8){$u$}
\put(92,93){$s$}
\put(72,100){$d$}
\put(59,103){$d$}
\put(90,73){$u$}
\put(89,49){$u$}
\put(158,48){+}
\put(35,-25){$6\, I_{12}(1^{uu}1^{uu}1^{uu}0^{us})\, \alpha_1^{0^{us}}$}
\put(180,54){\line(1,0){18}}
\put(180,53){\line(1,0){18}}
\put(180,52){\line(1,0){18}}
\put(180,51){\line(1,0){18}}
\put(180,50){\line(1,0){18}}
\put(180,49){\line(1,0){18}}
\put(180,48){\line(1,0){18}}
\put(180,47){\line(1,0){18}}
\put(180,46){\line(1,0){18}}
\put(210,50){\circle{25}}
\put(199,46){\line(1,1){15}}
\put(202,41){\line(1,1){17}}
\put(207.5,38.5){\line(1,1){14}}
\put(210,62){\vector(1,3){9}}
\put(224,84){9}
\put(196,35){\circle{16}}
\put(190,30){\vector(-3,-2){17}}
\put(170,24){6}
\put(190,30){\vector(-1,-4){5}}
\put(189,8){5}
\put(214,30){\circle{16}}
\put(215,22){\vector(1,-2){9}}
\put(228,2){3}
\put(215,22){\vector(-1,-2){9}}
\put(199,2){4}
\put(224,64){\circle{16}}
\put(229,44){\circle{16}}
\put(232,64){\vector(1,1){14}}
\put(240,80){7}
\put(232,64){\vector(3,-1){17}}
\put(242,63){1}
\put(237,46){\vector(1,1){12}}
\put(245,46){2}
\put(237,46){\vector(3,-2){14}}
\put(244,29){8}
\put(257.5,59){\circle{16}}
\put(265.5,59){\vector(3,2){16}}
\put(275,74){1}
\put(265.5,59){\vector(3,-2){16}}
\put(275,38){2}
\put(190,32){\small $1^{uu}$}
\put(208,27){\small $1^{uu}$}
\put(219,61){\small $0^{ud}$}
\put(223,41){\small $0^{ud}$}
\put(251.5,56){\small $1^{uu}$}
\put(287,74){$u$}
\put(287,41){$u$}
\put(229,-8){$u$}
\put(199,-8){$u$}
\put(183,-1){$u$}
\put(164,11){$u$}
\put(252,82){$d$}
\put(255,29){$d$}
\put(209,89){$s$}
\put(250,68){$u$}
\put(236,52){$u$}
\put(180,-25){$6\, I_{13}(1^{uu}1^{uu}0^{ud}0^{ud})\, \alpha_2^{0^{ud}0^{ud}}$}
\put(299,48){+}
\put(320,54){\line(1,0){18}}
\put(320,53){\line(1,0){18}}
\put(320,52){\line(1,0){18}}
\put(320,51){\line(1,0){18}}
\put(320,50){\line(1,0){18}}
\put(320,49){\line(1,0){18}}
\put(320,48){\line(1,0){18}}
\put(320,47){\line(1,0){18}}
\put(320,46){\line(1,0){18}}
\put(350,50){\circle{25}}
\put(339,46){\line(1,1){15}}
\put(342,41){\line(1,1){17}}
\put(347.5,38.5){\line(1,1){14}}
\put(350,62){\vector(1,3){9}}
\put(364,84){9}
\put(336,35){\circle{16}}
\put(330,30){\vector(-3,-2){17}}
\put(310,24){6}
\put(330,30){\vector(-1,-4){5}}
\put(329,8){5}
\put(354,30){\circle{16}}
\put(355,22){\vector(1,-2){9}}
\put(368,2){3}
\put(355,22){\vector(-1,-2){9}}
\put(339,2){4}
\put(364,64){\circle{16}}
\put(369,44){\circle{16}}
\put(372,64){\vector(1,1){14}}
\put(380,80){7}
\put(372,64){\vector(3,-1){17}}
\put(382,63){1}
\put(377,46){\vector(1,1){12}}
\put(385,46){2}
\put(377,46){\vector(3,-2){14}}
\put(384,29){8}
\put(397.5,59){\circle{16}}
\put(405.5,59){\vector(3,2){16}}
\put(415,74){1}
\put(405.5,59){\vector(3,-2){16}}
\put(415,38){2}
\put(330,32){\small $1^{uu}$}
\put(348,27){\small $1^{uu}$}
\put(359,61){\small $0^{ud}$}
\put(363,41){\small $0^{us}$}
\put(391.5,56){\small $1^{uu}$}
\put(427,74){$u$}
\put(427,41){$u$}
\put(369,-8){$u$}
\put(339,-8){$u$}
\put(323,-1){$u$}
\put(304,11){$u$}
\put(392,82){$d$}
\put(395,29){$s$}
\put(349,89){$d$}
\put(390,68){$u$}
\put(376,52){$u$}
\put(320,-25){$12\, I_{13}(1^{uu}1^{uu}0^{ud}0^{us})\, \alpha_2^{0^{ud}0^{us}}$}
\end{picture}

\vskip100pt

\begin{picture}(600,60)
\put(10,48){+}
\put(25,54){\line(1,0){18}}
\put(25,53){\line(1,0){18}}
\put(25,52){\line(1,0){18}}
\put(25,51){\line(1,0){18}}
\put(25,50){\line(1,0){18}}
\put(25,49){\line(1,0){18}}
\put(25,48){\line(1,0){18}}
\put(25,47){\line(1,0){18}}
\put(25,46){\line(1,0){18}}
\put(55,50){\circle{25}}
\put(44,46){\line(1,1){15}}
\put(47,41){\line(1,1){17}}
\put(52.5,38.5){\line(1,1){14}}
\put(50,61){\vector(1,3){10}}
\put(63,86){9}
\put(43,34){\circle{16}}
\put(40,26.5){\vector(1,-4){5.2}}
\put(48,3){5}
\put(40,26.5){\vector(-1,-1){15}}
\put(17,9){6}
\put(64,68){\circle{16}}
\put(64,32){\circle{16}}
\put(67,53){\vector(3,1){28}}
\put(88,51){2}
\put(67,47){\vector(3,-1){28}}
\put(88,42){4}
\put(72,70){\vector(3,2){21}}
\put(83,86){7}
\put(72,70){\vector(3,-1){23}}
\put(88,68){1}
\put(72,30){\vector(3,1){23}}
\put(88,25){3}
\put(72,30){\vector(3,-2){21}}
\put(83,7){8}
\put(103,60){\circle{16}}
\put(103,40){\circle{16}}
\put(111.5,60){\vector(3,2){21}}
\put(139,71){1}
\put(111.5,60){\vector(3,-1){23}}
\put(140,52){2}
\put(111.5,40){\vector(3,1){23}}
\put(140,42){3}
\put(111.5,40){\vector(3,-2){21}}
\put(139,19){4}
\put(37,31){\small $1^{uu}$}
\put(58,65){\small $0^{ud}$}
\put(58,29){\small $0^{ud}$}
\put(97,57){\small $1^{uu}$}
\put(97,37){\small $1^{uu}$}
\put(144,82){$u$}
\put(148,62){$u$}
\put(148,36){$u$}
\put(144,10){$u$}
\put(36,-1){$u$}
\put(22,0){$u$}
\put(96,89){$d$}
\put(96,8){$d$}
\put(51,93){$s$}
\put(77,60){$u$}
\put(74,49){$u$}
\put(81,44){$u$}
\put(76,35){$u$}
\put(20,-23){$24\, I_{14}(1^{uu}1^{uu}0^{ud}0^{ud})\, \alpha_2^{0^{ud}0^{ud}}$}
\put(156,48){+}
\put(175,54){\line(1,0){18}}
\put(175,53){\line(1,0){18}}
\put(175,52){\line(1,0){18}}
\put(175,51){\line(1,0){18}}
\put(175,50){\line(1,0){18}}
\put(175,49){\line(1,0){18}}
\put(175,48){\line(1,0){18}}
\put(175,47){\line(1,0){18}}
\put(175,46){\line(1,0){18}}
\put(205,50){\circle{25}}
\put(194,46){\line(1,1){15}}
\put(197,41){\line(1,1){17}}
\put(202.5,38.5){\line(1,1){14}}
\put(200,61){\vector(1,3){10}}
\put(213,86){9}
\put(193,34){\circle{16}}
\put(190,26.5){\vector(1,-4){5.2}}
\put(198,3){5}
\put(190,26.5){\vector(-1,-1){15}}
\put(167,9){6}
\put(214,68){\circle{16}}
\put(214,32){\circle{16}}
\put(217,53){\vector(3,1){28}}
\put(238,51){2}
\put(217,47){\vector(3,-1){28}}
\put(238,42){4}
\put(222,70){\vector(3,2){21}}
\put(233,86){7}
\put(222,70){\vector(3,-1){23}}
\put(238,68){1}
\put(222,30){\vector(3,1){23}}
\put(238,25){3}
\put(222,30){\vector(3,-2){21}}
\put(233,7){8}
\put(253,60){\circle{16}}
\put(253,40){\circle{16}}
\put(261.5,60){\vector(3,2){21}}
\put(289,71){1}
\put(261.5,60){\vector(3,-1){23}}
\put(290,52){2}
\put(261.5,40){\vector(3,1){23}}
\put(290,42){3}
\put(261.5,40){\vector(3,-2){21}}
\put(289,19){4}
\put(187,31){\small $1^{uu}$}
\put(208,65){\small $0^{ud}$}
\put(208,29){\small $0^{us}$}
\put(247,57){\small $1^{uu}$}
\put(247,37){\small $1^{uu}$}
\put(294,82){$u$}
\put(298,62){$u$}
\put(298,36){$u$}
\put(294,10){$u$}
\put(186,-1){$u$}
\put(172,0){$u$}
\put(246,89){$d$}
\put(246,8){$s$}
\put(201,93){$d$}
\put(227,60){$u$}
\put(224,49){$u$}
\put(231,44){$u$}
\put(226,35){$u$}
\put(170,-23){$48\, I_{14}(1^{uu}1^{uu}0^{ud}0^{us})\, \alpha_2^{0^{ud}0^{us}}$}
\put(306,48){+}
\put(325,54){\line(1,0){18}}
\put(325,53){\line(1,0){18}}
\put(325,52){\line(1,0){18}}
\put(325,51){\line(1,0){18}}
\put(325,50){\line(1,0){18}}
\put(325,49){\line(1,0){18}}
\put(325,48){\line(1,0){18}}
\put(325,47){\line(1,0){18}}
\put(325,46){\line(1,0){18}}
\put(355,50){\circle{25}}
\put(344,46){\line(1,1){15}}
\put(347,41){\line(1,1){17}}
\put(352.5,38.5){\line(1,1){14}}
\put(353,62){\vector(1,2){14}}
\put(372,87){8}
\put(350,61){\vector(1,4){7.5}}
\put(360,89){9}
\put(348,31){\circle{16}}
\put(344,24){\vector(1,-4){5.5}}
\put(354,0){5}
\put(344,24){\vector(-1,-1){16}}
\put(320,7){6}
\put(356,62.5){\vector(3,1){40}}
\put(385,80){1}
\put(374,57){\circle{16}}
\put(371,37){\circle{16}}
\put(381,61){\vector(1,1){15}}
\put(392,63){2}
\put(381,61){\vector(1,-1){12}}
\put(382,46){3}
\put(379,35){\vector(1,1){14}}
\put(383,32){4}
\put(379,35){\vector(1,-1){16}}
\put(398,11){7}
\put(404,77){\circle{16}}
\put(401,50){\circle{16}}
\put(412,78){\vector(3,2){18}}
\put(435,87){1}
\put(412,78){\vector(3,-2){18}}
\put(435,64){2}
\put(409,50){\vector(3,2){18}}
\put(431,54){3}
\put(409,50){\vector(3,-2){18}}
\put(431,32){4}
\put(342,28){\small $1^{uu}$}
\put(368,54){\small $1^{uu}$}
\put(365,34){\small $0^{ud}$}
\put(398,74){\small $1^{uu}$}
\put(395,47){\small $1^{uu}$}
\put(441,95){$u$}
\put(444,69){$u$}
\put(441,55){$u$}
\put(440,28){$u$}
\put(340,-6){$u$}
\put(322,-4){$u$}
\put(386,7){$d$}
\put(370,99){$d$}
\put(350,97){$s$}
\put(375,75){$u$}
\put(379,65){$u$}
\put(388,57){$u$}
\put(389,38){$u$}
\put(310,-23){$48\, I_{15}(1^{uu}1^{uu}1^{uu}0^{ud})\, \alpha_2^{1^{uu}0^{ud}}$}
\end{picture}

\vskip100pt

\begin{picture}(600,60)
\put(10,48){+}
\put(25,54){\line(1,0){18}}
\put(25,53){\line(1,0){18}}
\put(25,52){\line(1,0){18}}
\put(25,51){\line(1,0){18}}
\put(25,50){\line(1,0){18}}
\put(25,49){\line(1,0){18}}
\put(25,48){\line(1,0){18}}
\put(25,47){\line(1,0){18}}
\put(25,46){\line(1,0){18}}
\put(55,50){\circle{25}}
\put(44,46){\line(1,1){15}}
\put(47,41){\line(1,1){17}}
\put(52.5,38.5){\line(1,1){14}}
\put(53,62){\vector(1,2){14}}
\put(72,87){8}
\put(50,61){\vector(1,4){7.5}}
\put(60,89){9}
\put(48,31){\circle{16}}
\put(44,24){\vector(1,-4){5.5}}
\put(54,0){5}
\put(44,24){\vector(-1,-1){16}}
\put(20,7){6}
\put(56,62.5){\vector(3,1){40}}
\put(85,80){1}
\put(74,57){\circle{16}}
\put(71,37){\circle{16}}
\put(81,61){\vector(1,1){15}}
\put(92,63){2}
\put(81,61){\vector(1,-1){12}}
\put(82,46){3}
\put(79,35){\vector(1,1){14}}
\put(83,32){4}
\put(79,35){\vector(1,-1){16}}
\put(98,11){7}
\put(104,77){\circle{16}}
\put(101,50){\circle{16}}
\put(112,78){\vector(3,2){18}}
\put(135,87){1}
\put(112,78){\vector(3,-2){18}}
\put(135,64){2}
\put(109,50){\vector(3,2){18}}
\put(131,54){3}
\put(109,50){\vector(3,-2){18}}
\put(131,32){4}
\put(42,28){\small $1^{uu}$}
\put(68,54){\small $1^{uu}$}
\put(65,34){\small $0^{us}$}
\put(98,74){\small $1^{uu}$}
\put(95,47){\small $1^{uu}$}
\put(141,95){$u$}
\put(144,69){$u$}
\put(141,55){$u$}
\put(140,28){$u$}
\put(40,-6){$u$}
\put(22,-4){$u$}
\put(86,7){$s$}
\put(70,99){$d$}
\put(50,97){$d$}
\put(75,75){$u$}
\put(79,65){$u$}
\put(88,57){$u$}
\put(89,38){$u$}
\put(10,-25){$24\, I_{15}(1^{uu}1^{uu}1^{uu}0^{us})\, \alpha_2^{1^{uu}0^{us}}$}
\put(158,48){+}
\put(178,54){\line(1,0){18}}
\put(178,53){\line(1,0){18}}
\put(178,52){\line(1,0){18}}
\put(178,51){\line(1,0){18}}
\put(178,50){\line(1,0){18}}
\put(178,49){\line(1,0){18}}
\put(178,48){\line(1,0){18}}
\put(178,47){\line(1,0){18}}
\put(178,46){\line(1,0){18}}
\put(208,50){\circle{25}}
\put(197,46){\line(1,1){15}}
\put(200,41){\line(1,1){17}}
\put(205.5,38.5){\line(1,1){14}}
\put(203,61){\vector(1,3){10}}
\put(216,86){9}
\put(195,34){\circle{16}}
\put(193,26){\vector(1,-4){5.5}}
\put(203,2){5}
\put(193,26){\vector(-1,-1){16}}
\put(169,9){6}
\put(217,68){\circle{16}}
\put(217,32){\circle{16}}
\put(228,50){\circle{16}}
\put(225,70){\vector(3,2){21}}
\put(236,86){7}
\put(225,70){\vector(3,-1){23}}
\put(241,68){1}
\put(225,30){\vector(3,1){23}}
\put(241,25){4}
\put(225,30){\vector(3,-2){21}}
\put(236,7){8}
\put(236,50){\vector(1,1){12}}
\put(235,56){2}
\put(236,50){\vector(1,-1){12}}
\put(235,37){3}
\put(256,60){\circle{16}}
\put(256,40){\circle{16}}
\put(265,60){\vector(3,2){21}}
\put(292,71){1}
\put(265,60){\vector(3,-1){23}}
\put(293,52){2}
\put(265,40){\vector(3,1){23}}
\put(293,42){3}
\put(265,40){\vector(3,-2){21}}
\put(292,19){4}
\put(189,31){\small $1^{uu}$}
\put(211,65){\small $0^{ud}$}
\put(211,29){\small $0^{ud}$}
\put(222,47){\small $1^{uu}$}
\put(250,57){\small $1^{uu}$}
\put(250,37){\small $1^{uu}$}
\put(299,78){$u$}
\put(301,54){$u$}
\put(301,41){$u$}
\put(299,13){$u$}
\put(190,-4){$u$}
\put(177,0){$u$}
\put(247,89){$d$}
\put(247,6){$d$}
\put(203,89){$s$}
\put(234,69){$u$}
\put(243,51){$u$}
\put(244,44){$u$}
\put(233,27){$u$}
\put(150,-25){$24\, I_{16}(1^{uu}1^{uu}1^{uu}0^{ud}1^{uu}0^{ud})\, \alpha_3^{0^{ud}0^{ud}1^{uu}}$}
\put(312,48){+}
\put(328,54){\line(1,0){18}}
\put(328,53){\line(1,0){18}}
\put(328,52){\line(1,0){18}}
\put(328,51){\line(1,0){18}}
\put(328,50){\line(1,0){18}}
\put(328,49){\line(1,0){18}}
\put(328,48){\line(1,0){18}}
\put(328,47){\line(1,0){18}}
\put(328,46){\line(1,0){18}}
\put(358,50){\circle{25}}
\put(347,46){\line(1,1){15}}
\put(350,41){\line(1,1){17}}
\put(355.5,38.5){\line(1,1){14}}
\put(353,61){\vector(1,3){10}}
\put(366,86){9}
\put(345,34){\circle{16}}
\put(343,26){\vector(1,-4){5.5}}
\put(353,2){5}
\put(343,26){\vector(-1,-1){16}}
\put(319,9){6}
\put(367,68){\circle{16}}
\put(367,32){\circle{16}}
\put(378,50){\circle{16}}
\put(375,70){\vector(3,2){21}}
\put(386,86){7}
\put(375,70){\vector(3,-1){23}}
\put(391,68){1}
\put(375,30){\vector(3,1){23}}
\put(391,25){4}
\put(375,30){\vector(3,-2){21}}
\put(386,7){8}
\put(386,50){\vector(1,1){12}}
\put(385,56){2}
\put(386,50){\vector(1,-1){12}}
\put(385,37){3}
\put(406,60){\circle{16}}
\put(406,40){\circle{16}}
\put(415,60){\vector(3,2){21}}
\put(442,71){1}
\put(415,60){\vector(3,-1){23}}
\put(443,52){2}
\put(415,40){\vector(3,1){23}}
\put(443,42){3}
\put(415,40){\vector(3,-2){21}}
\put(442,19){4}
\put(339,31){\small $1^{uu}$}
\put(361,65){\small $0^{ud}$}
\put(361,29){\small $0^{us}$}
\put(372,47){\small $1^{uu}$}
\put(400,57){\small $1^{uu}$}
\put(400,37){\small $1^{uu}$}
\put(449,78){$u$}
\put(451,54){$u$}
\put(451,41){$u$}
\put(449,13){$u$}
\put(340,-4){$u$}
\put(327,0){$u$}
\put(397,89){$d$}
\put(397,6){$s$}
\put(353,89){$d$}
\put(384,69){$u$}
\put(393,51){$u$}
\put(394,44){$u$}
\put(383,27){$u$}
\put(335,-25){$48\, I_{16}(1^{uu}1^{uu}1^{uu}0^{ud}1^{uu}0^{us})\, \alpha_3^{1^{uu}0^{ud}0^{us}}$}
\end{picture}

\vskip110pt

\begin{picture}(600,100)
\put(10,48){+}
\put(25,54){\line(1,0){18}}
\put(25,53){\line(1,0){18}}
\put(25,52){\line(1,0){18}}
\put(25,51){\line(1,0){18}}
\put(25,50){\line(1,0){18}}
\put(25,49){\line(1,0){18}}
\put(25,48){\line(1,0){18}}
\put(25,47){\line(1,0){18}}
\put(25,46){\line(1,0){18}}
\put(55,50){\circle{25}}
\put(44,46){\line(1,1){15}}
\put(47,41){\line(1,1){17}}
\put(52.5,38.5){\line(1,1){14}}
\put(49,69){\circle{16}}
\put(46,76.5){\vector(1,4){5.5}}
\put(55,95){5}
\put(46,76.5){\vector(-1,1){16}}
\put(25,82){6}
\put(68,66){\circle{16}}
\put(74,71){\vector(1,1){16}}
\put(81,87){7}
\put(74,71){\vector(2,-1){18.5}}
\put(89,68){1}
\put(67,53){\vector(3,1){25.5}}
\put(81,50){2}
\put(101,62){\circle{16}}
\put(109,62){\vector(2,1){18}}
\put(122,75){1}
\put(109,62){\vector(2,-1){18}}
\put(122,43){2}
\put(53,30){\circle{16}}
\put(70,37){\circle{16}}
\put(75,30){\vector(2,-1){16}}
\put(90,27){8}
\put(75,30){\vector(-1,-4){4.2}}
\put(76,18){3}
\put(56,22.5){\vector(3,-2){14.5}}
\put(58,11){4}
\put(56,22.5){\vector(-1,-3){5.7}}
\put(44,8){9}
\put(74,5){\circle{16}}
\put(77,-2.5){\vector(3,-2){15}}
\put(93,-9){3}
\put(77,-2.5){\vector(-1,-3){5.7}}
\put(63,-18){4}
\put(43,66){\small $1^{uu}$}
\put(62,63){\small $0^{ud}$}
\put(95,59){\small $1^{uu}$}
\put(47,27){\small $0^{us}$}
\put(64,34){\small $0^{ud}$}
\put(68,2){\small $1^{uu}$}
\put(131,76){$u$}
\put(131,46){$u$}
\put(99,-17){$u$}
\put(77,-23){$u$}
\put(42,97){$u$}
\put(21,95){$u$}
\put(96,89){$d$}
\put(97,16){$d$}
\put(45,-4){$s$}
\put(82,71){$u$}
\put(73,48){$u$}
\put(68,22){$u$}
\put(61,21){$u$}
\put(-7,-40){$48\, I_{18}(1^{uu}1^{uu}1^{uu}0^{ud}0^{ud}0^{us})\, \alpha_3^{0^{ud}0^{ud}0^{us}}$}
\put(149,48){+}
\put(175,54){\line(1,0){18}}
\put(175,53){\line(1,0){18}}
\put(175,52){\line(1,0){18}}
\put(175,51){\line(1,0){18}}
\put(175,50){\line(1,0){18}}
\put(175,49){\line(1,0){18}}
\put(175,48){\line(1,0){18}}
\put(175,47){\line(1,0){18}}
\put(175,46){\line(1,0){18}}
\put(205,50){\circle{25}}
\put(194,46){\line(1,1){15}}
\put(197,41){\line(1,1){17}}
\put(202.5,38.5){\line(1,1){14}}
\put(199,69){\circle{16}}
\put(196,76.5){\vector(1,4){5.5}}
\put(205,95){5}
\put(196,76.5){\vector(-1,1){16}}
\put(175,82){6}
\put(218,66){\circle{16}}
\put(224,71){\vector(1,1){16}}
\put(231,87){7}
\put(224,71){\vector(2,-1){18.5}}
\put(239,68){1}
\put(217,53){\vector(3,1){25.5}}
\put(231,50){2}
\put(251,62){\circle{16}}
\put(259,62){\vector(2,1){18}}
\put(272,75){1}
\put(259,62){\vector(2,-1){18}}
\put(272,43){2}
\put(203,30){\circle{16}}
\put(220,37){\circle{16}}
\put(225,30){\vector(2,-1){16}}
\put(240,27){8}
\put(225,30){\vector(-1,-4){4.2}}
\put(226,18){3}
\put(206,22.5){\vector(3,-2){14.5}}
\put(208,11){4}
\put(206,22.5){\vector(-1,-3){5.7}}
\put(194,8){9}
\put(224,5){\circle{16}}
\put(227,-2.5){\vector(3,-2){15}}
\put(243,-9){3}
\put(227,-2.5){\vector(-1,-3){5.7}}
\put(213,-18){4}
\put(193,66){\small $1^{uu}$}
\put(212,63){\small $0^{us}$}
\put(245,59){\small $1^{uu}$}
\put(197,27){\small $0^{ud}$}
\put(214,34){\small $0^{ud}$}
\put(218,2){\small $1^{uu}$}
\put(281,76){$u$}
\put(281,46){$u$}
\put(249,-17){$u$}
\put(227,-23){$u$}
\put(192,97){$u$}
\put(171,95){$u$}
\put(246,89){$s$}
\put(247,16){$d$}
\put(195,-4){$d$}
\put(232,71){$u$}
\put(223,48){$u$}
\put(218,22){$u$}
\put(211,21){$u$}
\put(180,-40){$24\, I_{18}(1^{uu}1^{uu}1^{uu}0^{us}0^{ud}0^{ud})\, \alpha_3^{0^{ud}0^{ud}0^{us}}$}
\put(-7,-65){Fig. 1. The graphical equations of the reduced amplitude $\alpha_3^{1^{uu}1^{uu}1^{uu}}$
$I=2$ $ ^3_{\Sigma} He$.}
\end{picture}

\end{document}